\documentclass[11pt,twoside,final]{huthesis}

\usepackage{epsfig,bm,epsf,float,framed, soul}
\usepackage{graphicx,color,amsmath,bbm,makeidx}

\usepackage[english]{babel} 
\usepackage[latin1]{inputenc}
\usepackage[english]{nomencl}

\usepackage{amsmath,amssymb,amsfonts,amsthm,amsopn}
\usepackage{latexsym} 
\usepackage[T1]{fontenc}

\usepackage{setspace} 
\usepackage{import}
\usepackage{ctable}
\usepackage{afterpage}

\usepackage{verbatim}
\usepackage[]{subfigure} 
\usepackage{url} \usepackage{hyperref}

\begin{document}

%

\def\cal#1{\mathcal{#1}}

\def\avg#1{\left< #1 \right>}
\def\abs#1{\left| #1 \right|}
\def\recip#1{\frac{1}{#1}}
\def\vhat#1{\hat{{\bf #1}}}
\def\smallfrac#1#2{{\textstyle\frac{#1}{#2}}}
\def\smallrecip#1{\smallfrac{1}{#1}}

\def\spshalf{{1\over{2}}}
\def\btt#1{{\tt$\backslash$#1}}

\def\ylm{{Y_l^m}}
\def\b#1{\bmath#1} 
\def\Var{{\rm Var}} 
\def\expect#1{\left\langle #1\right\rangle}
\def\scale#1{\left[ #1\right]}
\def\T{\rm T_*}
\def\V{\rm V_*}
\def\vari#1{\Var\expect{#1}}
\def\I{{\mathbbmss{1}}}
\def\yr{{\rm\,yr}}
\def\st{\sin \theta}
\def\todo{$\heartsuit$}

\def\ts{\thinspace}
\def\gapprox{$_>\atop{^\sim}$}  
\def\lapprox{$_<\atop{^\sim}$}
\def\msun{{\rm\,M_\odot}}
\def\lsun{{\rm\,L_\odot}}
\newdimen\sa  \def\sd{\sa=.1em  \ifmmode $\rlap{.}$''$\kern -\sa$
                                \else \rlap{.}$''$\kern -\sa\fi}
              \def\se{\sa=.1em  \rlap{.}{''}\kern -\sa}
              \def\dgd{\sa=.1em \ifmmode $\rlap{.}$^\circ$\kern -\sa$
                                \else \rlap{.}$^\circ$\kern -\sa\fi}
\newdimen\sb  \def\md{\sa=.06em \ifmmode $\rlap{.}$'$\kern -\sa$
                                \else \rlap{.}$'$\kern -\sa\fi}

\def\msund{{\rm\,M_\odot}}
\def\lsund{{\rm\,L_\odot}}
\def\rsun{{\rm\,R_\odot}}

\def\sec{{\rm Sec.}}
\def\chapt{{\rm Chapter}}

\definecolor{light}{gray}{0.8} 
\definecolor{dark}{gray}{0.1}
\long\def\crap#1{}
\def\grommet{{\sc grommet}}
\def\ps{{\rm phase-space}}
\def\sch{{Schwarzschild}}
\def\tint{\int\!\!\int\!\!\int\!}

\def\halo{{\sf halo}}
\def\sub{{\sf sub}}
\def\bulge{{\sf bulge}}
\def\tbody{{\sf 2body}}
\def\modelb{{\sf bulge + MBH}}
\def\MBH{{\sf MBH}}
\def\subhalo{{\sf subhalo}}

\def\code{\rm {\bf code}}
\def\lp{\rm left panel}
\def\Lp{\rm Left panel}
\def\rp{\rm right panel}
\def\Rp{\rm Right panel}
\def\tp{\rm top panel}
\def\Tp{\rm Top panel}
\def\bp{\rm bottom panel}
\def\Bp{\rm Bottom panel}

\def\O{{\cal O}}
\def\d{{\rm d}}
\def\t{{\rm t}}
\def\E{{\cal E}}
\def\p{\partial}
\def\tint{\int\!\!\int\!\!\int\!}
\def\vt{{v_{\rm t}}}
\def\vr{{v_{\rm r}}}
\def\vp{{v_{\rm p}}}
\long\def\crap#1{}
\def\x{{\bmath x}} 
\def\v{{\bmath v}}
\def\a{{\bmath a}}
\def\t{{\bmath t}} 
\def\w{{\bmath w}}
\def\dw{\Delta{\bmath w}}
\def\dt{\Delta t}
\def\J{{\bmath J}}
\def\200{{{200}}}
\def\sgra{{\sf Sgr$A^*$}}
\def\c{{\rm c}}
\def\coll{{\rm coll}}
\def\df{{\rm df}}
\def\drain{^{\rm drain}}
\def\e{{\rm e}}
\def\ej{{\rm ej}}
\def\E{{\cal E}}
\def\h{{\rm h}}

\def\H{{\rm H}}
\def\Hubble{_{\rm Hubble}}
\def\lc{{\rm lc}}
\def\MF{{\rm MF}}
\def\esp{{\rm esp}}
\def\cir{{\rm cir}}
\def\gr{_{\rm gr}}
\def\lr{_{\rm lr}}
\def\lw{_{\rm lw}}
\def\max{{\rm max}}
\def\min{{\rm min}}
\def\peak{{\rm peak}}
\def\r{{\rm r}}
\def\s{{\rm s}}
\def\t{{\rm t}}
\def\tri{{\rm tri}}
\def\Var{{\rm Var}}
\def\expect#1{\left\langle #1\right\rangle}
\def\vari#1{\Var\expect{#1}}

\def\tm{\tilde M}
\def\tb{\tilde b}
\def\tv{\tilde V}
\def\te{\tilde \E}
\def\tj{\tilde J}
\def\tjj{\tilde J^2}

\def\np{{\sf N_p}}
\def\nbody{{$N$-body }}
\def\N{{$N$}}
\def\md{{\sf M_0}}
\def\rd{{\sf r_0}}
\def\vd{{\sf V_0}}
\def\td{{\sf T_0}}

\def\mbg{{\sf M_{bulge}}}
\def\rbg{{\sf r_{bulge}}}
\def\vbg{{\sf V_{bulge}}}
\def\tbg{{\sf T_{bulge}}}

\def\mha{{\sf M_{halo}}}
\def\rha{{\sf r_{halo}}}
\def\vha{{\sf V_{halo}}}
\def\tha{{\sf T_{halo}}}

\def\mf{{\sf M_1}}
\def\rf{{\sf r_1}}
\def\vf{{\sf V_1}}
\def\tf{{\sf T_1}}
\def\GN{{\sf G_N}}

\def\mnras{MNRAS}
\def\apj{ApJ}
\def\aj{AJ}
\def\aap{A\&A}
\def\apjl{ApJL}
\def\apjs{ApJS}
\def\nat{Nature}

\def\ms{{M_{\rm s}}}
\def\vs{{V_{\rm s}}}
\def\tr{{T_{\rm r}}}
\def\tz{{T_{\phi}}}
\def\omz{{\Omega_{\phi}}}
\def\omr{{\Omega_r}}
\def\omp{{\Omega_p}}
\def\oms{{\Omega_{\rm s}}}
\def\omc{{\Omega_{\rm c}}}
\def\rs{{r_{\rm s}}}
\def\rhos{\rho_{\rm s}}
\def\rc{{r_{\rm c}}}
\def\es{{\epsilon_{\rm s}}}
\def\ebh{{\epsilon_{\bullet}}}
\def\mbh{M_{\bullet}}
\def\rbh{r_{\bullet}}
\def\ubh{\mu_{\bullet}}
\def\am{{angular-momentum}}
\def\detj{\delta J}
\def\jlc{J_{\rm lc}}
\def\jjlc{J^2_{\rm lc}}
\def\jj{J^2}
\def\j{{\rm j}}
\def\i{{\rm i}}
\def\jc{J_c}
\def\jjc{J_c^2}
\def\re{r_{\E}}
\def\rh{r_{\rm infl}}
\def\ah{a_{\rm h}}
\def\rg{r_{\rm g}}
\def\rt{r_{\rm t}}

\def\oa#1{#1_{\rm O.A.}}
\def\oa#1{\bar #1}

\def\simlt{\mathrel{\hbox{\rlap{\hbox{\lower3pt\hbox{$\sim$}}}\hbox{$<$}}}}
\def\simgt{\mathrel{\hbox{\rlap{\hbox{\lower3pt\hbox{$\sim$}}}\hbox{$>$}}}}

\def\DX{\big\langle  \Delta X   \big\rangle}
\def\DXX{\big\langle \Delta X^2 \big\rangle}
\def\DXn{D \big [  \Delta X   \big]}
\def\DXXn{D \big [  \Delta X^2  \big]}
\def\DR{\big\langle  \Delta X^2   \big\rangle}
\def\DRn{D \big [  \Delta X^2   \big]}
\def\DRR{\big\langle (\Delta X^2)^2 \big\rangle}
\def\DRRn{D \big [  (\Delta X^2)^2   \big]}

\def\halff{{\textstyle{1\over2}}}

\def\st{\sin \theta}
\def\ct{\cos \theta}
\def\stt{\sin^2 \theta}
\def\ctt{\cos^2 \theta}
\def\sttt{\sin^3 \theta}
\def\cttt{\cos^3 \theta}
\def\lm{_{\rm lm}}
\def\plm{P_{\rm lm}}

\def\ext{{\rm ext}}
\def\lc{{\rm lc}}
\def\mlc{M_\lc}
\def\flc{F^\lc}
\def\rlc{r_\lc}
\def\dmlc{\dot{M}_\lc}
\def\tdmlc{\tilde \dmlc}

\def\flc{F^{\rm lc}}

\def\km{{\rm\,km}}
\def\meter{{\rm\,m}}
\def\kms{{\rm\,km\,s^{-1}}}
\def\Hz{{\rm Hz}}
\def\kpc{{\rm\,kpc}}
\def\Mpc{{\rm\,Mpc}}
\def\mpc{{\rm\,Mpc}}
\def\watt{{\rm J~s^{-1}}}
\def\wt{{\rm W}}
\def\M{{\sf\,M}}
\def\mstar{{\rm\,m_*}}
\def\rstar{{\rm\,r_*}}
\def\gal{{\rm\,gal}}
\def\pc{{\rm\,pc}}
\def\cm{{\rm\,cm}}
\def\m{{\rm\,m}}
\def\yr{{\rm\,yr}}
\def\Gyr{{\rm\,Gyr}}
\def\gyr{{\rm\,Gyr}}
\def\Myr{{\rm\,Myr}}
\def\myr{{\rm\,Myr}}
\def\au{{\rm\,AU}}
\def\gm{{\rm\,g}}
\def\kg{{\rm\,kg}}
\def\ergps{{\rm\,erg\,s}^{-1}}
\def\erg{{\rm\,erg}}
\def\K{{\rm\,K}}
\def\rms{{\caps rms}}

\def\ngc{{\rm NGC}}
\def\hst{{\em HST}}

\def\iso#1{\bar #1}

\def\be{\begin{equation}}
\def\ee{\end{equation}}
\def\bea{\begin{eqnarray}}
\def\eea{\end{eqnarray}}
\def\bean{\begin{mathletters}\begin{eqnarray}}
\def\eean{\end{eqnarray}\end{mathletters}}

\newcommand{\tbox}[1]{\mbox{\tiny #1}}
\newcommand{\half}{\mbox{\small $\frac{1}{2}$}}
\newcommand{\pit}{\mbox{\small $\frac{\pi}{2}$}}
\newcommand{\sfrac}[1]{\mbox{\small $\frac{1}{#1}$}}
\newcommand{\mbf}[1]{{\mathbf #1}}
\def\text{\tbox}

\newcommand{\mV}{{\mathsf{V}}}
\newcommand{\mL}{{\mathsf{L}}}
\newcommand{\mA}{{\mathsf{A}}}
\newcommand{\lB}{\lambda_{\tbox{B}}}  
\newcommand{\ofr}{{(\mbf{r})}}       
\newcommand{\ofs}{{(\mbf{s})}}       


\def\ie{{\rm i.e.\ }}
\def\eg{{\rm e.g.,\ }}
\newcommand{\etal}{{\rm et al.\ }}
\newcommand{\ibid}{{\it ibid.\ }}

\def\gap{\hspace{0.2in}}

%

\newcounter{eqletter}
\def\mathletters{%
\setcounter{eqletter}{0}%
\addtocounter{equation}{1}
\edef\curreqno{\arabic{equation}}
\edef\@currentlabel{\theequation}
\def\theequation{%
\addtocounter{eqletter}{1}\thechapter.\curreqno\alph{eqletter}%
}%
}
\def\endmathletters{\setcounter{equation}{\curreqno}}


%
%


\dsp


\title{Loss cone refilling by flyby encounters} 

\titleb{A numerical study of massive black holes in galactic centres}

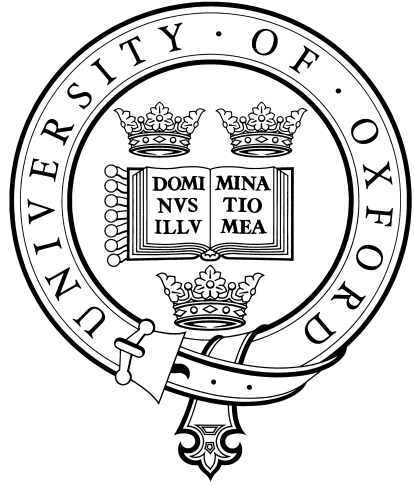
\author{Mimi Zhang}

\college{Wolfson College}
\univ{University of Oxford}

\degreemonth{Trinity Term} 
\degreeyear{2008}
\degree{Doctor of Philosophy} 
\field{Theoretical Astrophysics}
\department{Theoretical Physics}
\advisor{John S.  Magorrian} 

\maketitle
\copyrightpage

\begin{abstract}
\ssp
  
A gap in phase-space, the loss cone (LC), is opened up by a
supermassive black hole (MBH) as it disrupts or accretes stars in a
galactic centre.
  If a star enters the LC then, depending on its properties, its
  interaction with the MBH will either generate a luminous
  electromagnetic flare or give rise to gravitational radiation, both
  of which are expected to have directly observable consequences.
  A thorough understanding of loss-cone refilling mechanisms is
  important for the prediction of astrophysical quantities, such as
  rates of tidal disrupting main-sequence stars, rates of capturing
  compact stellar remnants and timescales of merging binary MBHs.
%
  If a galaxy were isolated and perfectly spherical, the only
  refilling mechanism would be diffusion due to weak two-body
  encounters between stars. This would leave the LC always nearly
  empty.
  However, real galaxies are neither perfectly spherical nor isolated.
  In this thesis, we use \nbody simulations to investigate how noise
  from accreted satellites and other substructures in a galaxy's halo
  can affect the LC refilling rate.

  Any \nbody model suffers from Poisson noise which is similar to, but
  much stronger than, the two-body diffusion occurring in real
  galaxies.
  To lessen this spurious Poisson noise, we apply the idea of
  importance sampling to develop a new scheme for constructing \nbody
  realizations of a galaxy model, in which interesting regions of
  phase-space are sampled by many low-mass particles.
  This scheme minimizes the mean-square formal errors of a given set
  of projections of the galaxy's phase-space distribution function.
  Tests show that the method works very well in practice, reducing the
  diffusion coefficients by a factor of $\sim~100$ compared to the
  standard equal-mass models and reducing the spurious LC flux in
  isolated model galaxies to manageably low levels.

  We use multimass \nbody models of galaxies with centrally-embedded
  MBHs to study the effects of satellite flybys on LC refilling rates.
  The total mass accreted by the MBH over the course of one flyby can
  be described, using a simple empirical fitting formula that depends
  on the satellite's mass and orbit.
  Published large-scale cosmological simulations yield predictions
  about the distribution of substructure in galaxy halos.  We use
  results of these together with our empirical fitting formula to
  obtain an upper bound on substructure-driven LC refilling rates in
  real galaxies.  We find that although the flux of stars into the
  initially emptied LC is enhanced, but the fuelling rate averaged
  over the entire subhalos is increased by only a factor 3 over the
  rate one expects from the Poisson noise due the discreteness of the
  stellar distribution.

  \vskip.1in

  \centerline{\it A thesis submitted for the degree of Doctor of
    Philosophy} 
  \centerline{\it Trinity Term 2008}

\end{abstract}


\newpage
\mbox{}

\addcontentsline{toc}{section}{Table of Contents}
\tableofcontents


\begin{acknowledgments}
  \ssp \vskip0.4in

  I wish to thank Dr. John Magorrian for providing me with a
  complicated jumbo holiday, for supporting me with constant
  entertainments, and for sharing with me the rewarding experience.
  My acknowledgments go to this fearless leader who draws me into
  galactic dynamics and gets me into \nbody simulations.
  No matter where I go, I will live with the courage he implants in
  me.
  All along the way, I have had good fortune to learn {\it Galactic
    Dynamics} with its author James Binney.

  \vskip.2in

  I owe a debt to Andy, who is instrumental in getting my programing
  started as a project.
  The ``Beecroft institute'' of Ralf, Rachel, Sarah, Andy, Michael,
  Ben, Callum have been great about sharing experience.
  And it is my duty to thank Dorothy Hodgkin Postgraduate PPARC-BP
  Awards for the financial support.

  \vskip.2in
  As seeming to languish for the last four years, I have challenged my
  willingness to see if I can lose my assumptions. 
  Replace my empty mind with an open one, and return with my empty
  mind in the end. What a failure and what an achievement!

  \vskip.2in
  I dedicate this thesis to my mother and father who cherish me more
  than I do, and vice versa.

  \vskip.5in

  \noindent{{\sf M$^2$Z}}\\
  \noindent{Oxford, June, 2008}
\end{acknowledgments}





\newpage

\startarabicpagination



\chapter{Introduction}
\label{chap:intro}

In the {\em high-redshift} universe, massive black holes (MBHs) have
been implicated as the powerhouses for quasars in the Active Galactic
Nuclei (AGN) paradigm;
in the {\em nearby} galaxies, the existence of dead quasar engines has
also been supported by modern stellar-dynamical searches (e.g.,
Ferrarese \& Ford 2005).
Over three dozens of local detections have unveiled the demographics
of MBH populations such as
a tight correlation between MBH {\em mass} and galactic {\em velocity
  dispersion} (e.g., Gebhardt \etal 2000; Ferrarese \& Merritt 2000;
Merritt \& Ferrarese 2001), and
a less tight correlation between MBH mass and the mass/luminosity of
the {\em bulge} (e.g. Kormendy \& Richstone 1995; Magorrian \etal
1998; Marconi \& Hunt 2003).
While efforts to build a larger and statistically significant sample
continue, it is time to understand the origin, evolution and cosmic
relevance of these fascinating objects.

The subject is large and my space is limited. In this thesis, I
concentrate on
developing and using \nbody models to learn more about the stellar
dynamics in galactic centres, especially around the MBHs.
I begin immediately below by reviewing the ``journey'' to the MBHs:
from the AGN paradigm-mandated MBHs as power sources to the
observation-supported MBHs.
Then, I describe two astrophysical ingredients: electromagnetic flares
by tidal disruption of main-sequence stars and gravitational wave
signals by capture of compact stars.
Next, I explain the emptiness of ``loss cone'' and the starvation of
the MBH which motivate the work in thesis.
Finally, I outline briefly \nbody modelling procedures developed in
the subsequent chapters.

\newpage

\section{Supermassive black holes in galactic centres}
\label{sec:MBHs}
 
\subsection{Energy arguments for MBHs in AGNs}
\label{sec:mbhagn}

\subsection{Loss of stars in the loss cone}
\label{sec:lcstars}
 
\subsection{How to feed stars to a loss cone?}
\label{sec:goals}
 
\subsubsection{Declaration}
\label{sec:declaration} 

\chapter{Dynamical modelling}
\label{chap:modelling}

Galaxies, as self-gravitating systems, can be idealized as
configurations of point mass fully described by a phase-space
distribution function (DF) $f(\x,\v,t)$.
The mass distribution determines the gravitational field[ through
Poisson's equation.
Over their lifetime, galaxies are to a high degree collisionless;
the DFs therefore satisfy the Collisionless Boltzmann equation
(\sec~\ref{sec:equilibrium}).
With the help of Monte Carlo methods (\sec~\ref{sec:MC}), the
evolution of the DF can be followed by \nbody integrations
(\sec~\ref{sec:CBE2nbody}).
When encounters are taken into account, 
one writes the collision term in Master equation form
(\sec~\ref{sec:gammaf}) and expands in a Taylor series to derive the
Fokker-Planck equation (\sec~\ref{sec:FK}).



\newpage

\section{Equilibrium model}
\label{sec:equilibrium}

\subsection{DFs for collisionless stellar system}
\label{sec:dfs}

\subsection{Collisionless Boltzmann equation}
\label{sec:CBE}

\subsection{Preliminaries}
\label{sec:modelref}

\subsubsection{Units conversion and scalings}
\label{sec:units}

\section{\nbody modelling}
\label{sec:nbody} 

\subsection{Monte Carlo integration}
\label{sec:MC}

\subsection{The CBE and \nbody simulations}
\label{sec:CBE2nbody}

\subsubsection{I.\,\,\,Initial conditions}
\label{sec:ics}

\subsubsection{II.\,\,\,Poisson solvers}
\label{sec:poissonsolvers}

\subsubsection{III.\,\,\,Leap-frog integrator}
\label{sec:leapfrog}

\subsubsection{Summary}

\section{Departure from equilibrium}
\label{sec:nonequilibrium}

\subsection{Collision terms}
\label{sec:gammaf}

\subsection{Fokker-Planck equation}
\label{sec:FK}

\subsection{Relaxation time}
\label{sec:tr}

\chapter[Multi-mass schemes for collisionless simulations]
{Multi-mass schemes \\[.1in]
  {\Large for collisionless \nbody simulations}}
\label{chap:multimass}

A general scheme for constructing Monte Carlo realizations of
equilibrium, collisionless galaxy models with known distribution
function (DF) $f_0$ is established.  It uses importance sampling to
find the sampling DF $f_s$ that minimizes the mean-square formal
errors in a given set of projections of the DF $f_0$.  The result is a
multi-mass \nbody realization of the galaxy model in which
``interesting'' regions of phase-space are densely populated by lots
of low-mass particles, increasing the effective $N$ there, and less
interesting regions by fewer, higher-mass particles.

The chapter is organized as follows. 
After recapping in the connection between \nbody simulations and the
CBE (\sec~\ref{sec:nbodyCBE}) and presenting two possible weapons to
fight against small \N\, limitation (\sec~\ref{sec:nbodyweapon}), %
we explain our multi-mass formulation in \sec~\ref{sec:basics}.
In \sec~\ref{sec:model1} we give two examples of using our scheme to
suppress fluctuations in the monopole component of acceleration in
spherical \bulge\, models with or without a central \MBH.
We calculate formal estimates of the noise in \nbody models
constructed using equal-mass scheme and SHQ95's method, and compare
them to our own scheme (\sec~\ref{sec:test1}).
In \sec~\ref{sec:test2} and \ref{sec:model2} we test how well our
realizations behave in practice when evolved using a real \nbody code.
For the \modelb\,model in \sec~\ref{sec:model2}, much care has been
taken to resolve the dynamics around the \MBH.

 \newpage
\section{Collisionless \nbody simulations}
\label{sec:cns}
 
\subsection{\nbody simulations and the CBE}
\label{sec:nbodyCBE}
 
\subsection{How to fight against small \N\,limitations?}
\label{sec:nbodyweapon}

\section{Formulation}
\label{sec:basics}
 
\subsection{Observables}
\label{sec:windowfn}
 
\subsection{Optimal sampling scheme}
\label{sec:choosefs}
 
\subsection{ICs for \nbody model}
\label{sec:ics}

\section{Comparisons of a \bulge\, model (no \MBH)}
\label{sec:model1}
 
\subsection{Formal errors}
\label{sec:comparison}
 
\subsubsection{I.\,\,The conventional equal-mass scheme}
\label{sec:fsequal}
 
\subsubsection{II.\,\,Sigurdsson \etal 's multi-mass scheme}
\label{sec:fssig}
 
\subsubsection{III.\,\,Our scheme}
\label{sec:fsmulti}
 
\subsubsection{\nbody realizations}
\label{sec:realizations1}
 
\subsection{How well is the acceleration field reproduced?}
\label{sec:test1} 
 
\subsection{How well are integrals of motion conserved?}
\label{sec:test2} 
 
\section{Simulations of a \modelb\, model}
\label{sec:model2}
 
\subsection{\nbody realizations}
\label{sec:realizations2}
 
\subsubsection{II.\,\,Block leapfrog time-stepping}
\label{sec:blockleapfrog}

\subsection{Results: diffusion in $J^2$}
\label{sec:results2}

\chapter{Loss cone refilling by flyby encounters}
\label{chap:lcrefilling}

In the last chapter, we have developed a general scheme for
constructing multi-mass collisionless galaxy models. 
It has been successfully applied to set up a \modelb\,model, whose
long-time behavior has also been verified by full \nbody experiments.
In this chapter, the MBH-embedded model is extended to include a loss
cone (LC): a consumption sphere centred on the MBH.
First of all, control tests have been directed to calibrate
(artificial) noise-driven flux into the LC.
Then starting from this near-equilibrium galaxy, satellites on flyby
orbits are added as transient perturbers.
For satellites with different mass and orbital parameters, I carry out
extensive experiments to measure the mass of stars captured into the
LC after a single flyby encounter.
Given a set of satellite parameters, an empirical relationship is
found to predict the LC refilling mass.
Finally, to understand the interactions between the perturber and
stars, especially to disentangle effects of resonance coupling from
velocity impulse, I run extra test-particle experiments of orbiting
satellites.

\crap{We take particular care to disentangle effects of Poisson noise
  from the genuine features in the galaxy's response to the applied
  perturbations.
  An empirical relationship is found to predict the captured mass given
  a set of satellite parameters.
  Finally, we deliberately construct test-particle experiments to track
  dynamics pertaining to resonances for an orbiting satellite,
  with the aim of understanding the response of stellar orbits to
  external perturbations.
}

\newpage

\section{Loss cone refilling by two-body relaxation}
\label{sec:lctheory}

\subsubsection{What is the size of a loss cone?}
\label{sec:geometry}
 
\subsubsection{How to refill a loss cone in spherical galaxies?}
\label{sec:lc2body}
 
\subsubsection{How to model a loss cone in \nbody simulations? }
\label{sec:modellc}

\section{Flyby encounters}
\label{sec:flyby}
 
\subsection{Tests: \ calibrating noise-driven LC refilling}
\label{sec:nbody2body}
 
\subsection{Experiments: \ measuring perturber-driven LC refilling}
\label{sec:nbodysate}
 
\subsubsection{I.\,\,Toy perturbers}
\label{sec:modeltoy}

\subsubsection{II.\,\,An example of flyby encounters}
\label{sec:asample}
 
\subsection{Results: \ an empirical $\mlc$ formula}
\label{sec:mlcfitting}
 
\subsubsection{The validity of $\mlc$ formula}
\label{sec:validity}
 
\subsection{Discussion}
\label{sec:discussion3}

\section{Where do LC stars come from?}
\label{sec:mlcinterp}

\subsection{Angular-momentum transport by satellite torque}
\label{sec:impulse}
 
\subsection{Resonances from perturbation theory}
\label{sec:resonance}
 
\subsection[Test-particle experiments of a periodic perturber] %
{Are resonances important?  \newline Test-particle experiments of a
  periodic perturber}
\label{sec:testptle}

\clearpage
\clearpage
\chapter{Rates of MBH feeding in bulge-halo systems}
\label{chap:feeding}

In this chapter, I study the prospects for feeding the central MBH by
orbiting halo substructures.
I start by reviewing substructure properties observed in
high-resolution cosmological \nbody simulations, 
then summarize information on the subhalo spatial distribution, radial
density profile, mass function and concentration parameters into a DF
of the form $f_\sub(\E; M)$.
In \sec~\ref{sec:flc}, I use the subhalo DF together with the
knowledge of how much mass can be fed to a MBH after a single
encounter to estimate the perturber-driven LC refilling rate for a
typical small galaxy.

\newpage

\crap{
  In \sec~\ref{sec:modelmw}, I review the observed Milky Way galaxy.
  Ignoring the disk component, our Galaxy is mainly made up of a central
  MBH, a stellar bulge and a clumpy dark matter halo.
  To probe the halo structure on $~\sim 100\pc$ scales, I resort to
  high-resolution cosmological \nbody simulations
  (\sec~\ref{sec:modelhalo}).
  There, I summarize the information on the subhalo mass function,
  spatial distribution and velocity field into a DF.
  I use subhalos' concentration parameters to determine its boundary
  conditions.
  Finally in \sec~\ref{sec:flc}, I estimated the rate LC repopulation
  by averaging over many single events, using the subhalos' DF.
}

\section{Modelling a Milky-way sized halo and its substructures}
\label{sec:modelmw}

\section{LC flux driven by noisy DM clumps}
\label{sec:flc}

\section{Discussions}
\label{sec:discuss5}

\clearpage
\clearpage
\chapter{Conclusions and future studies}
\label{chap:conclusions}

The motivation behind this thesis work was to improve our
understanding of the internal dynamics at galactic centres, 
in particular, around super massive black holes.
The best available tool on the dynamical study is undoubtedly full
self-consistent \nbody simulations.
We have (1) \,\, developed a multi-mass scheme for constructing
collisionless \nbody models and (2)\,\, used the multi-mass \modelb\,
model to study loss cone dynamics.
Below, I summarize results in earlier chapters and suggest some future
work.  \newpage

\section{Constructing collisionless galaxy models}

In \chapt~\ref{chap:modelling}, I presented the basic dynamical
equations that govern the evolution of collisionless systems.
Beginning with the collisionless Boltzmann equation that is of
essential importance for system (\sec~\ref{sec:CBE}),
I show explicitly how the CBE is related to the \nbody modelling
(\sec~\ref{sec:nbody}).
When encounters are taken into account, one writes the collision term
in Master equation form and expands in a Taylor series to derive the
Fokker-Planck equation (\sec~\ref{sec:nonequilibrium}).

Rather technical, Chapter~\ref{chap:multimass} forms the backbone of
the thesis. In it we saw %
how to design a sampling distribution function (DF) $f_s$ from some
known DF $f_0$ (\sec~\ref{sec:choosefs}) and 
how to generate a multi-mass collisionless model from it
(\sec~\ref{sec:ics}).
In two applications, a \bulge\, model and a \modelb\, model, we aim at
minimizing the shot noise in estimates of the acceleration field.
Models constructed using our multi-mass scheme easily yield a factor
$\sim100$ reduction in the variance at the central acceleration field
when compared to a traditional equal-mass model with the same number
of particles.  When evolving both models with an \nbody code, the
diffusion coefficients in our model are reduced by a similar factor.
Therefore, for certain types of problems, our scheme is a practical
method for reducing the two-body relaxation effects, thereby bringing
the \nbody simulations closer to the collisionless ideal. 
We note the following preparation for the applications of multi-mass
modelling scheme:
\begin{itemize}

\item For successful application, a system should be
  in a steady state, or close to one.

\item The DF $f_0$ should be quick and cheap to evaluate, either
  numerically or analytically.  
  Finding $f_0$ for axisymmetric or triaxial galaxies is a
  longstanding and nontrivial problem since one rarely has sufficient
  knowledge of the underlying potential's integrals of motion, but
  suitable flattened DFs do exist, including the standard axisymmetric
  two-integral $f(\E,L_z)$ models and also rotating triaxial models
  such as those used in, \eg Berczik \etal (2006) An alternative way
  of constructing flattened multi-mass realizations would be to apply
  Holley-Bockelmann \etal (2002)'s adiabatic sculpting scheme to a
  spherical \nbody model constructed using our scheme.

\item The general multi-mass scheme uses importance sampling to find
  the tailored sampling DF $f_s$ that minimizes the sum of mean-square
  uncertainties in $Q_i$ (of the form eq.~\ref{eq:proj}).
  As long as $f_s$ is smooth in integral space, Monte Carlo
  realizations of $f_s$ should work for any reasonably general
  collisionless \nbody code.
  The utility of our multi-mass scheme, therefore, depends critically
  on the selection of the projection kernels $Q_i(\w)$.

\end{itemize}
\vskip.3cm
The last point is new to this field.  It is probably best addressed by
experimenting with different sets of kernels, especially since it is
easy to test the consequences of modifying them.  Nevertheless, there
are cases in which modest physical insight offers some guidance on
choosing the $Q_i$.  Besides the loss cone problems addressed in
\chapt~\ref{chap:lcrefilling}, I here give another example for {\bf
  future work}:
\begin{enumerate}

  \crap{
  \item[\todo 0.] Loss-cone problems\qquad The rate of supplying stars
    into a MBH's loss-cone is an important ingredient in galaxy models
    with central MBHs. A thorough understanding of collisionless
    loss-cone refilling mechanisms and accurate estimates of the
    resulting refilling rates are particularly critical for the
    prediction of astrophysical quantities such as the timescale of
    binary MBH merger (Begelman \etal 1980; Yu 2002; Milosavljevi{\'c}
    \& Merritt 2003b), the tidal disruption rate
    of stars (Syer \& Ulmer 1999; MT99; Wang \& Merritt 2004). When
    using \nbody simulations to study such loss-cone problems, one is
    often interested in {\em stars on low angular momentum orbits} and
    can therefore choose kernels to pick out such loss-cone phase-space
    for detailed modelling, while simultaneously maintaining accurate
    estimates of the galaxy's acceleration field.  }

\item[\todo 1.] Sinking satellites\qquad Kazantzidis \etal (2004)
  demonstrate the significance of using equilibrium \nbody
  realizations of satellite models when investigating the effect of
  tidal stripping of DM substructure halos (satellites) orbiting
  inside a more massive host potential.  Besides the shape of the
  background potential and the amount of tidal heating, the mass-loss
  history is very sensitive to the detailed density profile of the
  satellite itself.  One can therefore make one step further from
  equal-mass realizations by designing kernels to pick out {\em orbits
    that pass through the tidal radius}, while again maintaining an
  accurate estimate of the satellite's acceleration field.
\end{enumerate}

\section{Rates of loss cone refilling by MBHs}

\chapt~\ref{chap:lcrefilling} describes applications of a multi-mass
\modelb\, model to study the loss cone (LC) dynamics.
In \sec~\ref{sec:lctheory}, I introduced a loss cone consumption
sphere to mimic tidal disruption events.
The ``empty'' nature of the LC was then scrutinized by calibrating
artificial LC refilling (\sec~\ref{sec:nbody2body}).
Toy models in \sec~\ref{sec:modeltoy} taught us that only a fairly
massive satellite $(\ms)$ passing through the centre $(b)$ with low
speed $(V)$ can give rise to a LC flux well above the threshold of
numerical noise-driven LC refilling.
From the reduced satellite parameter space, about 100 experiments have
been carried out
to measure the maximum net mass $(\mlc)$ captured into an initially
empty LC.
The collected data were {\bf summarized} by a fitting
$\mlc$-formula~(\ref{eq:mlc-fit-all-2b}) in \sec~\ref{sec:mlcfitting}.
\begin{itemize}

\item Stellar orbits become strongly perturbed during a satellite
  flyby.
  The LC-captured mass reaches a peak value soon after the perturber
  visits its pericentre.

\item $\mlc$ is strongly correlated with the satellite parameters:
  scaling linearly with the satellite mass ($\ms$) and loss cone size
  ($\rlc$), but inversely with $b^{2.0}$ and $V^{1.3}$.

\item As the satellite departs, stars settle down to a new equilibrium
  configuration within a few crossing times.
  Therefore, if perturbed by cumulative perturbers, one can simply
  accumulate each single event to get the total consumed
  mass. 
\end{itemize}
The rest of the \chapt~\ref{chap:lcrefilling} is devoted to
understanding numerical results: where do LC stars come from?
Linear perturbation theory claims that resonances are important if not
most important in determining the response of a stellar system to any
external perturbation (\sec~\ref{sec:resonance}).
But for the flyby encounters of interest, could resonances be the
culprit for redistributing \am\,in stars and eventually send stars to
the LC?
This was addressed in \sec~\ref{sec:testptle}, where test-particle
experiments were carried out to track resonant dynamics relevant to a
periodic driver.
I make the following {\bf summaries}:
\begin{itemize}

  \crap{
  \item Sufficiently close to the MBH, tightly bound stars are {\bf
      adiabatically} protected from wandering into the LC sphere.
    During the flyby event, ``changes that occur in the structure of the
    orbits as the perturber approaches will be reversed as it departs,
    and the encounter will leave most orbits in the central region
    unchanged.''  (page 658, BT08).

  \item Away from the MBH, where the crossing time within the stars is
    large, the impulse approximation takes over adiabatic protection.
    Close encounters with the satellite effectively change \am\, of
    stars. The ones that lose enough \am\, can be effectively deflected
    into the loss-cone and consumed by the MBH.

  \item This is because a flyby encounter does not repeat its pattern
    and prevents stars in instantaneous resonance from building up a
    significant response.
  }

\item Close encounters with the satellite effectively change \am\, of
  stars. The ones that lose enough \am\, can be effectively deflected
  into the loss cone and destroyed by falling into the MBH.

\item   A flyby encounter does not repeat its pattern, which prevents stars
  in instantaneous resonance from building up a significant response.
  Therefore, resonance coupling between stars and the external
  perturber is unimportant in terms of refilling the emptied loss
  cone.
\item The accelerated \am\, relaxation within a nearly Keplerian is
  also unimportant here.

\end{itemize}
In \chapt~\ref{chap:feeding}, I introduced an outer halo as a
reservoir to continuously inject orbiting subhalos into the inner
bulge.
One further step was made by describing the subhalo properties in
terms of phase-space probability density, containing all the dynamical
information (\sec~\ref{sec:modelmw}).
Finally in \sec~\ref{sec:flc}, I put things together to predict a LC
refilling rate.
The conclusion is:

\begin{itemize}
\item The flux of stars into the loss cone is {\bf enhanced} when the
  loss cone is initially emptied, but due to the scarcity of subhalo
  population near the galactic centre,
  the LC refilling rate averaged over the entire orbiting dark halo
  substructures is {\bf not} strongly affected.
\end{itemize}
\subsubsection{The future}
For the galaxy model being used in the entire thesis, I have
introduced several simplifications.
At this point, I bring some notes about possible improvements and
extensions of the current model.

\begin{enumerate}
\item[\todo 2.] Better galaxy models \qquad The galaxy model used in
  this thesis is not ideal:
  ({\sf a}) \,\, the mass of the MBH is considerably larger than those
  observed in galactic centres and the effective radius ($\rlc$) of
  the LC is many orders of magnitude larger than the tidal radius in
  real galaxies. 
  Unfortunately, it is still hard for simulations to determine
  accurately the flux of stars into a small MBH or a tiny LC. The only
  remedy is to increase the number of particles, the more the better.
  ({\sf b}) \,\, The adopted Hernquist profile has a $r^{-1}$ central cusp
  and no alternatives, such as core (shallower) or nucleus (steeper)
  profiles, have been explored here.
  It would be useful to extend the work in this thesis by building
  \nbody models with different mass profiles, MBH masses and LC sizes.

\item[\todo 3.] Evolution of binary supermassive black holes \qquad
  Larger elliptical galaxies and bulges grow through mergers. If more
  than one of the progenitors contains a MBH and the in-spiral time is
  less than a Hubble time, the MBHs will form a bound
  system. Begelman, Blandford \& Rees (1980) showed that the evolution
  of binary MBHs in gas-poor galaxies can be divided into three
  phases:\,\, 
  ({\sf a}) As galaxies merge, the core undergoes violent relaxation.
  Via dynamical friction (Chandrasekhar 1943), the captured MBH sinks
  towards the center of the common gravitational potential where they
  form a bound binary. \,\,
  ({\sf b}) The binary continues to decay via gravitational slingshot
  interactions (Saslaw, Valtonen \& Aarseth 1974): stars on orbits
  intersecting the binary are ejected at much higher velocities
  comparable to the binary's orbital velocity.  \,\,
  ({\sf c}) Finally \textbf{if} the binary's separation decreases to
  the point at which gravitational radiation becomes the dominant
  dissipative force to carry away the last remaining energy and
  angular momentum, the binaries coalesce rapidly (Peters 1964).
The long-term evolution of MBH binaries is always related to a long
standing problem called the ``final parsec problem''
(Milosavljevi\'{c} \& Merritt 2003b): is super-elastic scattering
off individual stars in the background efficient enough to transit a
binary MBH from dynamical friction regime on the order of $1 {\rm
  pc}$ scale, to the gravitational radiation dominated regime with a
separation $\leq 10^{-3} {\rm pc}$ and finally merge within a Hubble
time?
Uncertainties about the resolution have been a major impediment to
predicting the frequency of MBH mergers in galactic nuclei, and
hence to computing event rates for proposed gravitational wave
interferometers like LISA.\footnote{http://lisa.jpl.nasa.gov/}
As a natural continuation of the work in \chapt~\ref{chap:feeding},
I intend to further explore the use of $\mlc$-fitting-formula for
the evolution of binary MBHs.

\item[\todo 4.] Response of galaxy to perturbations \qquad Our
  conclusion that resonances are {\bf unimportant} in redistributing
  \am\, within stars, is in disagreement with Vesperini \& Weinberg
  (2000), who explore the effects of relatively weak encounters by
  computing the response of a spherical stellar system to the
  perturbation induced by a low-mass system during a flyby.
  They determine the contribution to the total self-consistent
  response from three sources: ({\sf a}) the perturbation applied by
  the external perturber; ({\sf b}) the reaction of the system to its
  own response to this perturbation; ({\sf c}) the excitation of
  discrete damped modes of the primary system;
and conclude that {\em it is the resonances} (point c) {\em that
  lead to the excitation of patterns in the primary system.}
The main discrepancies probably stem from the galaxy model under
study, since all their conclusions are based on a very strong
assumption: discrete modes are weakly damped in {\em King} models
with different concentrations.
On the theoretical side, I intend to explore the response of galaxy
to external perturbations by means of linear perturbation theory,
hopefully to find out whether weakly-damped modes exist in every
galaxy.
\end{enumerate}


\ssp
\nocite{}


\appendix
\include{ap1}
\include{ap2}


\end{document}